\documentclass[sigconf]{acmart}
\usepackage{multirow}
\usepackage{booktabs}
\usepackage{balance}

\newcommand\blfootnote[1]{%
  \begingroup
  \renewcommand\thefootnote{*}\footnotetext{#1}%
  \addtocounter{footnote}{-1}%
  \endgroup
}

\AtBeginDocument{%
  \providecommand\BibTeX{{%
    \normalfont B\kern-0.5em{\scshape i\kern-0.25em b}\kern-0.8em\TeX}}}

\begin{document}

\title[Modeling Relevance Ranking under the Pre-training and Fine-tuning Paradigm]{Modeling Relevance Ranking  \\under the Pre-training and Fine-tuning Paradigm}
\author{Lin Bo,\textsuperscript{1} Liang Pang,\textsuperscript{3} Gang Wang,\textsuperscript{4} Jun Xu,$^{2,*}$ XiuQiang He,\textsuperscript{4} Ji-Rong Wen\textsuperscript{2}}
\affiliation{%
  \institution{\textsuperscript{1}School of Information, Renmin University of China}
  \institution{\textsuperscript{2}Gaoling School of Artificial Intelligence, Renmin University of China}
  \institution{\textsuperscript{3}Institute of Computing Technology, Chinese Academy of Sciences; \textsuperscript{4}Huawei Noah's Ark Lab}
  \country{}
}
\email{{bolin20, junxu, jrwen}@ruc.edu.cn, pangliang@ict.ac.cn, {wanggang110,hexiuqiang1}@huawei.com}



\begin{abstract}\blfootnote{Corresponding author}
Recently, pre-trained language models such as BERT have been applied to document ranking for information retrieval (IR). These methods usually first pre-train a general language model on an unlabeled large corpus and then conduct ranking-specific fine-tuning on expert-labeled relevance datasets. Though preliminary successes have been observed in a variety of IR tasks, a lot of room still remains for further improvement. Ideally, an IR system would model relevance from a user-system dualism: the user's view and the system's view. User's view judges the relevance based on the activities of ``real users'' while the system's view focuses on the relevance signals from the system side, e.g., from the experts or algorithms, etc.~\cite{saracevic1975relevance, hjorland2010foundation}. Inspired by the user-system relevance views and the success of pre-trained language models, in this paper we propose a novel ranking framework called Pre-Rank that takes both user's view and system's view into consideration, under the pre-training and fine-tuning paradigm. Specifically, to model the user's view of relevance, Pre-Rank pre-trains the initial query-document representations based on a large-scale user activities data such as the click log. To model the system's view of relevance, Pre-Rank further fine-tunes the model on expert-labeled relevance data. More importantly, the pre-trained representations, are fine-tuned together with handcrafted learning-to-rank features under a wide and deep network architecture. In this way, Pre-Rank can model the relevance by incorporating the relevant knowledge and signals from both real search users and the IR experts. 
To verify the effectiveness of Pre-Rank, we showed two implementations by using BERT~\cite{devlin2018bert} and SetRank~\cite{pang2020setrank} as the underlying ranking model, respectively. Experimental results base on three publicly available benchmarks showed that in both of the implementations, Pre-Rank can respectively outperform the underlying ranking models and achieved state-of-the-art performances. The results demonstrate the effectiveness of Pre-Rank in combining the user-system views of relevance. 

\end{abstract}


\keywords{pre-trained IR model; neural information retrieval}

\maketitle

\section{Introduction}
Relevance ranking, whose objective is to provide the right ranking order of a list of documents for a given query~\cite{liu2011learning, li2011learning}, has played a vital role in the field of information retrieval (IR). Machine learning models, especially deep neural networks~\cite{goodfellow2016deep} have been applied to relevance ranking and many ranking techniques have been developed~\cite{guo2020deep, xu2018deep}. One branch of the research formalizes the learning of ranking models as first pre-training a general language model on large-scale unlabeled texts and then fine-tuning on the labeled relevance data. 
%For example, \citet{nogueira2019passage} propose to XXX. In \citet{ICTXXX}, the authors propose a new loss function to training the IR model 
For example, \citet{nogueira2019passage} consider the training of a passage ranking model as a downstream task in BERT fine-tuning; \citet{chang2020pre} propose to pre-train a BERT using the Inverse Cloze Task (ICT) as the objective, which aims to teach the model to predict the removed sentence given a context text; \citet{ma2021prop} proposes a new task of representative words prediction (ROP) to pre-train a BERT model for the ad-hoc retrieval. 

Despite improvements that have been observed in many ranking tasks, existing studies either directly consider the ranking learning as a downstream task under the pre-training framework~\cite{nogueira2019passage}, or simply adapt pre-training objectives. The nature and requirements of relevance ranking are rarely taken into consideration. According to the studies in~\cite{saracevic1975relevance,hjorland2010foundation}, the relevance of a document to a query can be considered from two views: the user's view and the system's view. The user's view prefers that the relevance assessments should be made by the ``real users''. The system's view also called the algorithm's view, emphasizes systems processing information objects and matching them with queries. Ideally, an IR model would consider both of these two views when conducting the relevance ranking of documents. Existing studies, however, usually model the relevance from only one of the views. 

Inspired by the observations and the recent progress of the pre-trained language models, in this paper we propose modeling both the user's and the system's view of relevance under the pre-training and fine-tuning framework, called Pre-Rank.
%Pre-Rank follows the pre-training and fine-tuning paradigm in the learning of the ranking model. 
In the pre-training stage, it makes use of the real user's activities, i.e., the large-scale user click log, rather than the unlabeled text corpus for training. Since the user activities imply the relevance judgments of the query-document pairs from the real users, the pre-trained representations are based on the user's view of relevance, rather than the general natural language processing (NLP) knowledge. One problem with the pre-trained representations is that the user's activities are usually noisy and contain strong biases (e.g., position bias, selection bias, etc.). To alleviate the issue, Pre-Rank fine-tunes the pre-trained model parameters with query-document pairs with unbiased labels by the experts. To further enhance the ranking performances and incorporating the expert knowledge, Pre-Rank also extends a wide branch~\cite{cheng2016wide} to the pre-trained network, resulting in a wide and deep architecture where the wide branch is responsible to inject the handcrafted learning-to-rank features into the model. Since the labeled query-document pairs and the handcrafted features were created based on the expert's knowledge on the relevance, the fine-tuning stage naturally adapts the ranking model to reflect the system's view on the relevance and alleviate the biases from the users. 

Pre-Rank offers several advantages. First of all, compared to existing studies in which the pre-trained models are based on unsupervised text data, Pre-Rank makes use of the click log. The pre-trained representations, therefore, reflect the user's view on relevance, which makes it easy to conduct the downstream fine-tune on expert-labeled relevance data. Second, the handcrafted learning-to-rank features and expert-labeled query-document pairs reflect the expert knowledge (system's view) on IR relevance, which is complementary to relevance information from user activities in the pre-training stage. It is believed that modeling the two views simultaneously is helpful, especially when very limited labeled query-document pairs are available for fine-tuning. Third, existing studies have shown that the user clicks contain biased information (e.g., position bias, selection bias, etc.) and therefore may hurt the model performances if being directly used as the training corpus ~\cite{chen2020bias}. Pre-Rank provides an effective pre-training and fine-tuning approach to relieving the bias issue.

Pre-Rank is a general framework that can involve several types of neural ranking models as its underlying ranking model. As examples, we present two implementations of Pre-Rank based on the state-of-the-art deep ranking models of BERT and SetRank. In the Pre-Rank implementation with BERT, we start with BERT$_{\textrm{BASE}}$ and continue the pre-training with binary cross-entropy loss on the click log. In the fine-tuning stage, the extended wide and deep BERT network is fine-tuned with expert-labeled data. In the implementation with SetRank, the model is pre-trained on the click log with list-wise cross-entropy loss. After extending with handcrafted features, the wide and deep SetRank network is fine-tuned with the labeled relevance data.

We conducted experiments to test the effectiveness of Pre-Rank by pre-training on large-scale ORCAS click log~\cite{craswell2020orcas}, and fine-tuning on three publicly available benchmarks of MQ2007, MQ2008~\cite{qin2010letor}, and TREC19~\cite{craswell2020overview}. In our experiments, we found that Pre-Rank can respectively outperform the underlying neural ranking models of BERT and SetRank, and achieved state-of-the-art ranking performances on three benchmarks. The results indicate the effectiveness of introducing click-data and handcrafted features in the pre-training and fine-tuning. The experimental analysis also showed that combining the relevance signals from the user's view and from the system's view can help to improve the accuracy of relevance ranking.

\section{RELATED WORK}
One of the most fundamental problems in information retrieval is how to interpret the concept of relevance. \citet{saracevic1975relevance, hjorland2010foundation} present ``the user's view'' and ``the system's view'' of relevance. The user's view formalize and measure relevance based on ``real users''. In practice, commercial search engines have collected large-scale user activity data (e.g., click log) which provide implicit relevance assessments from the real users. One benefit of using click log is it ``removes the cost to the experts of examining and rating the items.''~\cite{fox2005evaluating}. A large number of studies have been conducted to use the click log in search. For example, ~\citet{joachims2002optimizing} trains the ranking SVM model based on the document preference pairs constructed based on the user clicks. ~\citet{craswell2007random} apply a Markov random walk model to a large click log. Relevant documents that have not yet been clicked for that query can be retrieved and ranked effectively. \citet{agichtein2006improving} construct ranking features based on user feedback. See also~\cite{dou2008click}. One difficulty of using the activities from real users is that the log data contains much noise and biases. \citet{joachims2017accurately} examined the reliability of the implicit relevance information derived from the click data and found that the clicks are informative but biased, including the position biases, selection bias, etc. Therefore, the user clicks could not be treated as relevant judgments~
\cite{joachims2017accurately,radlinski2005query,joachims2007evaluating}. Directly using the click log will inherent biases which is a key difficulty to effetely use it~\cite{joachims2017unbiased}. Recently, a large number of methods (e.g., counterfactual learning) have been proposed to remove the bias information in the click log~\cite{yuan2020unbiased, yuan2019improving}.

In contrast, the system's view of relevance (also called algorithmic relevance) ``describes the relationship between the query (terms) and the collection of information objects expressed by the retrieved information object(s)''~\cite{borlund2003concept}. As one of the representative approaches, traditional learning-to-rank models~\cite{liu2011learning, li2011learning} represent the query-document pairs with handcrafted features by the experts, and learn the ranking models based on labeled relevance data by the annotators~\cite{ burges2005learning, burges2010ranknet, yue2007support, xu2007adarank}. One difficulty of the approach is the high cost of gathering high quality handcrafted features and relevant labels. In recent years, with the availability of large-scale datasets, deep neural networks have also been applied to IR ranking, called neural information retrieval~\cite{mitra2018introduction}. The neural information retrieval models can automatically learn the query-document features from the raw data, which bridge the gap between query and document vocabulary. For example, some deep approaches \cite{huang2013learning,guo2016deep,pang2016text} focus on discovering the interactions between query and documents and learn complicated interaction patterns. Both traditional handcrafted features and automatically learned features are crucial for relevance ranking~\cite{guo2020detext}.

More recently, pre-trained language representation models have led to significant improvements on many NLP tasks~\cite{devlin2018bert,radford2018improving,yang2019xlnet}. These models are pre-trained on a great amount of unlabeled data with a large neural architecture. As one of the most representative works, BERT~\cite{devlin2018bert} gets the language representation model by pre-training on large-scale unlabeled data on a bidirectional transformer-based model. By adding a simple feedforward classification layer on top of BERT, it can outperform many task-specific architectures on various tasks, including IR ranking. There have been studies \cite{nogueira2019passage,nogueira2020document,dai2019deeper,10.1007/978-3-030-72240-1_11} applied the pre-trained models to the search tasks, by feeding query-document pair into BERT and compute the relevance score over the multi-layer perception (MLP) layer of ``[CLS]'' token. State-of-the-art performances have been achieved. 
% 加上计算所的工作，对pre-trained部分的改进
In \cite{ma2021prop}, the authors designed a specific pre-training task for IR, called representative words prediction, by sampling a pair of word sets according to the document language model.

The pre-trained models usually employ plain text to learn the initial representations, which are originally designed from NLP tasks and do not match well with the task of relevance ranking. In this paper, we try to use a large-scale user click log to pre-train the initial representations and then fine-tune the model on expert-labeled data. From the user-system view of relevance, the work can be explained as a study of combining these two views in one model. See also the experimental settings in ~\cite{dai2019deeper,qiao2019understanding}.

\section{Pre-Rank: Our Approach}
\begin{figure}
    \centering
    \includegraphics[width=0.5\textwidth]{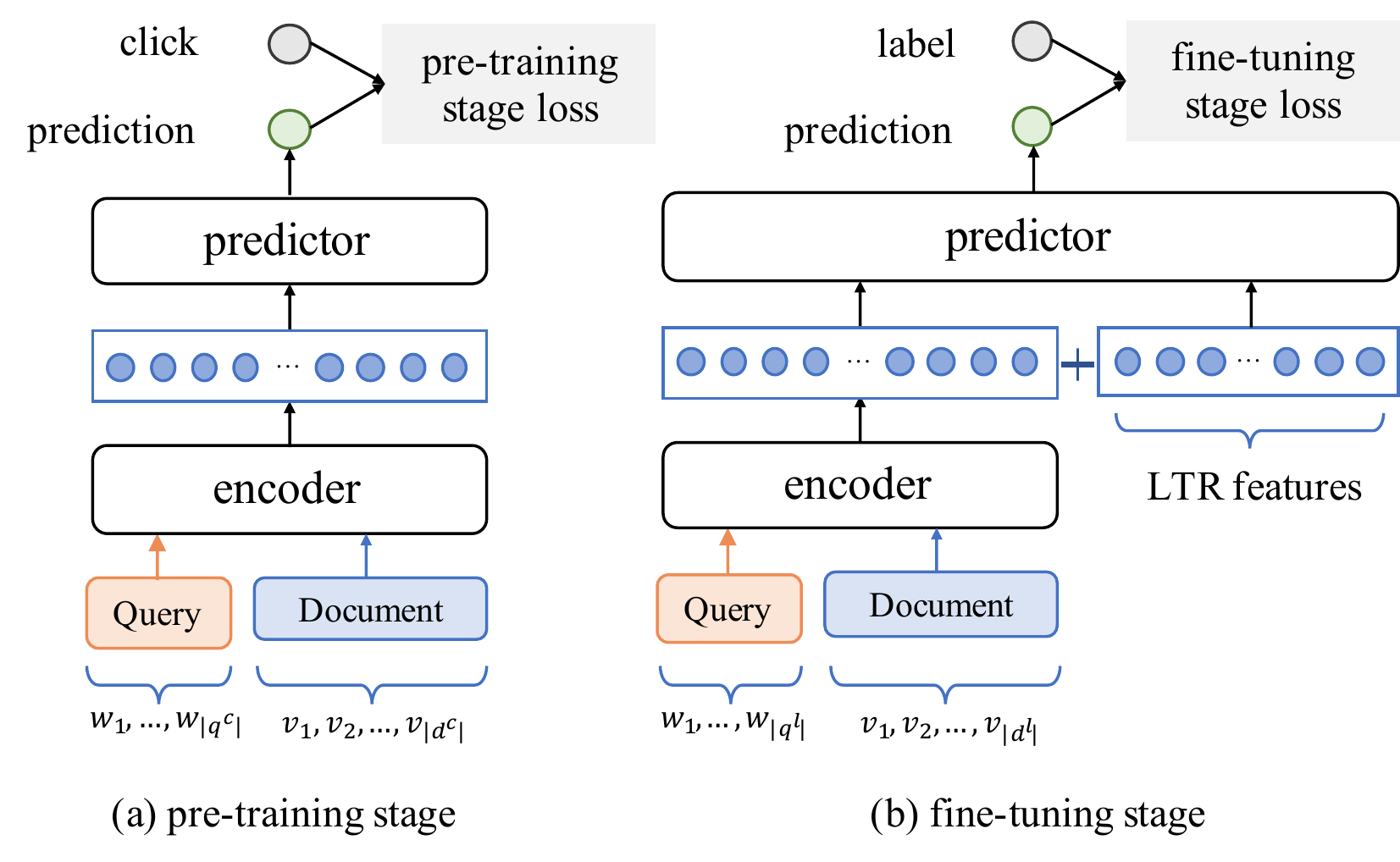}
    \caption{The pre-training (a) and fine-tuning (b) paradigm of Pre-Rank.}\label{fig:arch}
    \label{fig:overall_architecture}
\end{figure}

The proposed framework, called Pre-Rank, consists of two stages: pre-training stage and fine-tuning stage, in which the model learns the relevance from user's and system's view respectively. The overall structure can be seen in Figure ~\ref{fig:arch}.

The goal of the pre-training stage is to make the model incorporate specific knowledge of the search domain. In this stage, we used click log to generate supervision signals, because those data can be massively and cheaply obtained, and can reflect the implicit relevance from the user's view, although such type of signal has some biases problem. We considered raw text feature in this stage because it is a common feature across different ranking datasets.

With a good initialization of the model, the fine-tuning stage aims to directly optimize the target task. Different from the pre-training stage, the fine-tuning stage is suitable to consider dataset dependent features and use expert-labeled corpus as supervision signals. Though it is costly and time-consuming to construct large-scale handcrafted features and expert-labeled corpus, the clean relevance signals make them valuable to adjust the biases and relief the issues of the pre-trained models. Specifically, we used the combination of the learned features and handcrafted features to represent the query-document pairs and fine-tune the extended ranking model on the expert-labeled corpus.

\subsection{Modeling User's View of Relevance with Pre-train}
In the pre-training stage, it is supposed that we are given a set of search activities from the real users (e.g., large-scale click log): 
\begin{equation}
    \label{click_data}
    \mathbb{D}^c = \left\{(q^c,\mathbf{d}^c = (d_1^c,d_2^c\cdots,d_m^c), \mathbf{c} = (c_1,c_2\cdots,c_m))\right\},
\end{equation}
where $q^c=\left\{w_1,\cdots,w_{|q^c|}\right\}$ is a query inputted by real users, where $w_i$ denotes the $i$-th word in the query. Given the query $q^c$, an IR system will retrieve a set of documents $\mathbf{d}^c$ presented to the users, where $d_i^c$ denotes the $i$-th document in $\mathbf{d}^c$. $d_i^c=\left\{v_1,\cdots,v_{|d_i^c|}\right\}$ also consists of a list of words and $v_j$ denotes the $j$-th word in the document. Let $\mathbf{c}$ be the list of click signals associated with $\mathbf{d}^c$ where $c_i\in\{0, 1\}$ denotes whether the $i$-th document is clicked by the user where $c_i = 1$ if the user clicked $d^c_i$ and 0 otherwise. 

The pre-training stage aims to learn a ranking model $M^{pre}$ using a pre-training algorithm $\mathcal{A}^{pre}$, based on the click log $\mathbb{D}^c$:
$$
    M^{pre} \leftarrow \mathcal{A}^{pre}(\mathbb{D}^c). 
$$
Usually, the model $M^{pre}$ 
%.The overall structure of pre-trained stage can be seen in the left part of Figure~\ref{fig:overall_architecture}.
contains an encoder layer $E^{pre}(\cdot)$ and a prediction layer $P^{pre}(\cdot)$. The encoder layer aims to generate a vector that encodes the interactions between query and documents. The prediction layer aims to predict the possibility of user clicks on the retrieved documents given the query.

The training algorithm $\mathcal{A}^{pre}$ consists of two steps: first optimizes the masked language modeling (MLM) objective and then optimizes the click prediction objective. The MLM objective, proposed in~\cite{taylor1953cloze} for NLP, aims to build the contextual representations for queries and documents, the loss denotes as $\mathcal{L}_{MLM}$. Specifically, MLM masks out some tokens from input and then trains the model to predict the masked ones from the rest of the sentence. MLM has been proven to build good contextual representations in many NLP tasks. Following the existing practices~\cite{devlin2018bert}, the MLM loss is defined as
% \begin{equation}\label{eq:lossMLM}
%     \mathcal{L}_{MLM}=-\sum_{(q^c, \mathbf{d}^c, \mathbf{c})\in \mathbb{D}^c}\sum_{d_i^c\in\mathbf{d}^c; \hat{w}\in m( d_i^c)} \log p\left(\hat{w} \mid  d_i^c_{\backslash \hat{w}}\right),
% \end{equation}
\begin{equation}\label{eq:lossMLM}
\mathcal{L}_{MLM}=-\sum_{\hat{x} \in m(\mathbf{x})} \log p\left(\hat{x} \mid \mathbf{x}_{\backslash m(\mathbf{x})}\right),
\end{equation}
where $\mathbf{x}$ is the input sentences, $m(\mathbf{x})$ are the randomly masked words from $\mathbf{x}$,  $\mathbf{x}_{\backslash m(\mathbf{x})}$ represent the rest of words from $\mathbf{x}$, and $p$ is the prediction probability of the masked word $\hat{x}$~\cite{devlin2018bert}. 
% where $m(d_i^c)$ are the randomly masked words from $d_i^c$, and $d_i^c_{\backslash \hat{w}}$ represent the rest of words from $d_i^c$, and $p$ is the prediction probability of the masked word $\hat{w}$~\cite{devlin2018bert}. 

The click prediction objective models the probability of whether a user will click on the document from $\textbf{d}^c$ when entering query $q^c$, it reflects the relevance between $q^c$ and $\textbf{d}^c$ from the user's perspective. The loss denotes as $\mathcal{L}_{click}$ aims to minimize the differences between the click predictions and the real user clicks:
\begin{equation}\label{click_loss}
\mathcal{L}_{click}= \sum_{(q^c, \mathbf{d}^c, \mathbf{c})\in \mathbb{D}^c} \ell^{pre}(P^{pre}(E^{pre}(q^c,\mathbf{d}^c)),\mathbf{c}),
\end{equation}
where $\ell^{pre}$ is the query-level loss at pre-training stage.

Through optimizing the pre-training objectives, the pre-trained model $M^{pre}$ is able to reflect the implicit relevance between query and documents as well as the contextual interaction between query and documents. Noisy and large-scale are two major properties in this stage. Click-through data contains a lot of noise for the sake of positional bias and explosion bias. The noise in the inputs and supervision signals makes it hard to model the relevance, while large-scale data make it possible to involve relevance related information in the model.  %Also note that not all terms in the text contribute equally (e.g. stopwords, terms with small IDF values) and some terms do not affect relevance much. Thus, there exists a lot of noise if we use raw text as input. 
In the following fine-tuning stage, the expert-labeled data are used to alleviate these issues.

\subsection{Modeling System’s View of Relevance with Fine-tune}
In the fine-tuning stage, we are given a set of expert-labeled data $\mathbb{D}^l$ as the training corpus:
\begin{equation}
\label{label_dataset}
    \mathbb{D}^l = \left\{(q^l,\mathbf{d}^l = (d_1^l,d_2^l\cdots,d_m^l),\mathbf{l}=(l_1,l_2,\cdots,l_m))\right\},
\end{equation}
where $q^l$ is the given query and $\textbf{d}^l$ is the document list associate with query $q^l$, where $d_i^l$ is the $i$-th document in $\mathbf{d}^l$. Let $\mathbf{l}$ be the list of relevance labels associated with $\mathbf{d}^l$ where $l_i\in\{0, 1, \cdots, k\}$ denotes how the $i$-th document relevance to the query judge by expert and $k$ varies on different dataset. Moreover, $q^l=\left\{w_1,\cdots,w_{|q^l|}\right\}$,$d_i^l=\left\{v_1,\cdots,v_{|d_i^l|}\right\}$ representing respectively as the raw text of query and document, $w_i$ and $v_i$ denotes the $i$-th word in $q^l$ and $d_i^l$. 

During the fine-tuning stage, we aim to fine-tune the model $M^{pre}$ using algorithm $\mathcal{A}^{fine}$ with the expert-labeled data $\mathbb{D}^{l}$:
$$
    M^{fine} \leftarrow \mathcal{A}^{fine}(\mathbb{D}^l, M^{pre}).
$$
Similar to the pre-training model, $M^{fine}$ also contains an encoder layer $E^{fine}(\cdot)$ and a prediction layer $P^{fine}(\cdot)$. The encoder layer $E^{fine}$ shares the same structure with $E^{pre}$ and is initialized by the pre-trained parameters in $E^{pre}$. During fine-tuning, the parameters in $E^{fine}$ will be further tuned on the labeled data, takes $q^l$ and $\textbf{d}^l$ as input and the output is denoted as $\mathbf{h}^{fine}$, $\mathbf{h}^{fine} = E^{fine}(q^l, \textbf{d}^l)$. As for $P^{fine}$, the network structure is an extension of $P^{pre}$, by adding a wide branch for involving the handcrafted learning-to-rank features. The parameters in $P^{fine}$ are randomly initialized and tuned in the fine-tuning stage.   

The training algorithm $\mathcal{A}^{fine}$ aims to minimize the loss between prediction and human label $\mathbf{l}$, to get the final ranking list for labeled data:
\begin{equation}\label{label_loss}
\mathcal{L}_{label}= \sum_{(q^l, \mathbf{d}^l, \mathbf{l})\in \mathbb{D}^l} \ell^{fine}\left(P^{fine}\left([\mathbf{h}^{fine},\psi(q^l,\mathbf{d}^l)]\right),\mathbf{l}\right),
\end{equation}
where $\ell^{fine}$ is the query-level loss at the fine-tuning stage, $\psi(q^l, \mathbf{d}^l)$ outputs the handcrafted learning-to-rank features for the inputted query and document, and $[\cdot, \cdot]$ denotes the concatenation of the inputted vectors.  

During the fine-tuning stage, we use unbiased datasets labeled by experts that model relevance from the system's view, which can relieves the noise and bias issues with the click log. Clean but small-scale are the main properties in this stage.
We use the original features in the benchmark datasets if it is provided, if not, we handcraft ourselves.
With the training algorithm $\mathcal{A}^{fine}$, model $M^{fine}$ can predict the explicit relevance between query and documents and also leverage the handcrafted features.

\begin{figure*}
    \centering
    \includegraphics[width=\textwidth]{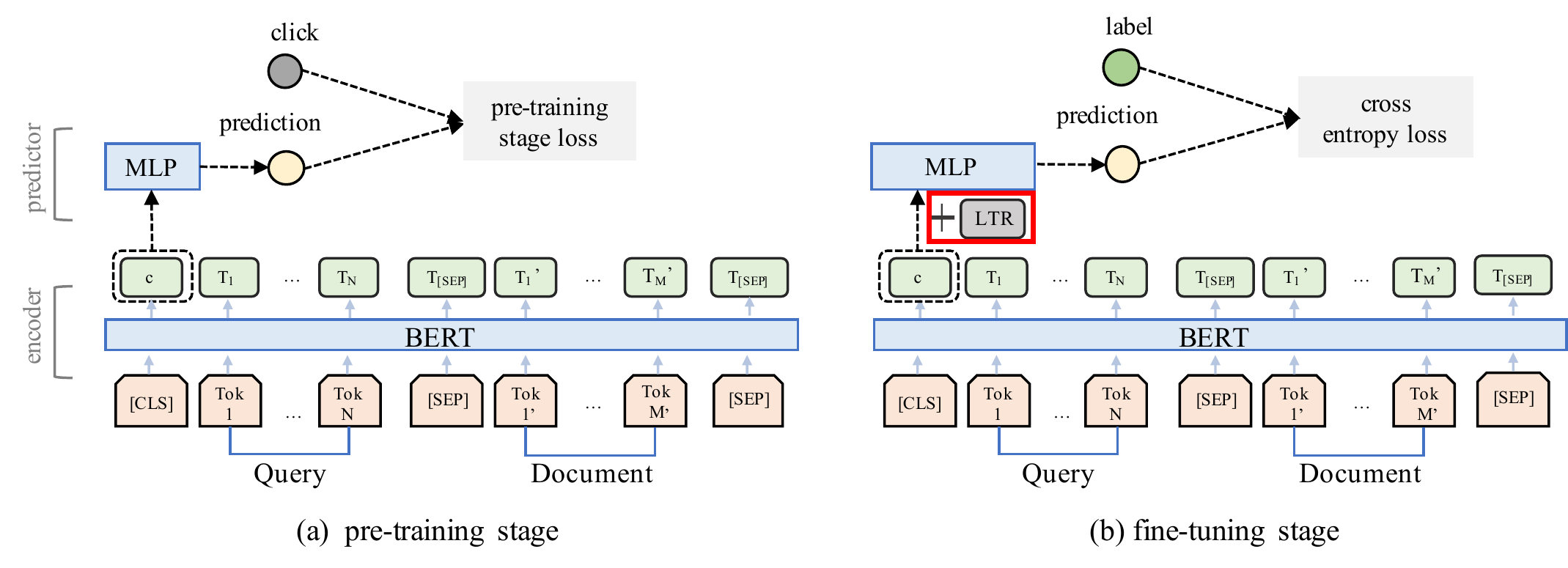}
    \caption{Implementation of Pre-Rank with BERT as the underlying model.}
    \label{fig:BERT_imp}
\end{figure*}

\section{Implementations}

Pre-Rank is a general ranking framework. In principle, it can be used to improve different ranking methods, by using the method as its underlying ranking model. In this section, we penetrate into the details of two Pre-Rank implementations which use BERT and SetRank as the underlying ranker, denote as Pre-Rank (BERT) and Pre-Rank (SetRank) respectively.
\subsection{Implementation with BERT}
%In this section, we first introduce BERT and then introduce the details of implementation.
BERT is a language representation model proposed by~\citet{devlin2018bert}. It pre-trains deep bidirectional representations on BooksCorpus~\cite{zhu2015aligning} and English Wikipedia by using the pre-training objectives of masked language model (MLM) and next sentence prediction. The first token of every input sequence to BERT is always a special classification token ([CLS]). 
To fine-tune on downstream tasks, the final hidden vector corresponding to the first input token ([CLS]) is feed into a classification layer to get the final prediction. In this paper, we use the off-the-shelf BERT to conduct the pre-training and fine-tuning. The overall architecture of implementing Pre-Rank with BERT is shown in Figure~\ref{fig:BERT_imp}. 

\subsubsection{Pre-training Stage}
The pre-training stage is shown in Figure~\ref{fig:BERT_imp}(a). During this stage, the model takes the click log as input. Since BERT is a pair inputted model, for adaptation, we use $(q_i^c,d_i^c,c_i) \in \mathbb{D}^c$ as one training instance.
% convert the training set $\mathbb{D}^c$ to $\mathbb{D}^c_{BERT} = \left\{(q_i^c,d_i^c,c_i)\right\}_{i=1}^N$, where $c_i = 1$ means that $d_i^c$ was clicked after the user issuing query $q_i$, and 0 otherwise, and $N$ is the number of records (query-document-click triples) in the click log. 
As for the encoder layer $E^{pre}$ in BERT, the input is the concatenation of query tokens and the clicked document tokens, with special delimiting tokens i.e.,
$
[CLS]+q_i^c+[SEP]+d_i^c+[SEP].
$
The tokens go through several layers of transformers to get fully interaction between query and document. Finally, the output embedding vector of the [CLS] token, denoted as $\mathbf{h}^{pre}_{i}$, is used as a representation for the entire query-document pair, which is obtained by 
$$
\mathbf{h}^{pre}_{i}=E^{pre}([CLS]+q_i^c+[SEP]+d_i^c+[SEP]).
$$
$\mathbf{h}^{pre}_{i}$ is then feed into the prediction layer $P^{pre}$, which is a multi-layer perception (MLP) to predict the possibility of click. Cross-entropy loss is used here as the learning objective. Therefore, Eq.~
(\ref{click_loss}) can be written as
\begin{equation}\label{eq:loss_click_ce}
\mathcal{L}_{click}= \sum_{(q^c_i, d_i^c, c_i)\in \mathbb{D}^c_{BERT}} \mathrm{CE}(P^{pre}(E^{pre}(q_i^c,d_i^c)),c_i),
\end{equation}
where CE denotes the cross-entropy loss function. 

During the pre-training stage, the model is first initialized with the pre-trained BERT$_{\textrm{BASE}}$'s parameters to leverage the language model, while the prediction layer is learned from scratch. The pre-training stage first optimizes the mask language model loss in Eq.~(\ref{eq:lossMLM}), and thereafter optimizes the click loss in Eq.~(\ref{eq:loss_click_ce}). 

\subsubsection{Fine-tuning Stage}
The fine-tuning stage is shown in Figure~\ref{fig:BERT_imp}(b). Similar to that of in the pre-training stage, we use $(q_i^l,d_i^l,l_i)\in\mathbb{D}^l$ as one training instance.
%we convert the fine-tune training set $\mathbb{D}^l$ to $\mathbb{D}^l_{BERT} = \left\{q_i^l,d_i^l,l_i\right\}_{i=1}^M$, where $l_i$ is the relevance label for $q_i^l$ and $d_i^l$, and $M$ is the number of labeled query-document pairs. 
%\liang{need modify}
As for the encoder $E^{fine}$ in $M^{fine}$, we initialize the parameters with the pre-trained encoder $E^{pre}$ from first stage, and fine-tune it with labeled data. Following the pre-training stage, the input is the concatenated query and documents tokens: $[CLS]+q_i^l+[SEP]+d_i^l+[SEP]$. The tokens go through several layers of transformers to get fully interaction, and achieve the representation corresponding to the [CLS] token:
$$
\mathbf{h}^{fine}_{i}=E^{fine}([CLS]+q_i^l+[SEP]+d_i^l+[SEP]).
$$

The output embedding of the [CLS] token $\mathbf{h}^{fine}_{i}$ is concatenated with the handcrafted learning-to-rank features $\psi(q_i^l,d_i^l)$,
and then feed into the predictor layer to predict the possibility of relevance. Please note that to involve the handcrafted features, the predictor becomes a wide and deep model architecture and the extended wide part corresponds to the learning-to-rank features. To learning the parameters, the cross-entropy loss between predictions and human labels is constructed and optimized. 
\begin{equation}
\mathcal{L}_{label}= \sum_{(q^l_i, d_i^l, l_i)\in \mathbb{D}^l_{BERT}} \mathrm{CE}\left(P^{fine}\left([\mathbf{h}^{fine}_{i}, \psi(q_i^l,d_i^l)]\right),l_i\right).
\end{equation}
 
\subsection{Implementation with SetRank}
In this section, we first introduce SetRank and then describe the details of implementation.

SetRank is a listwise ranking model described by \citet{pang2020setrank}. It is a permutation-invariant ranking model defined on document sets of any size. The architecture of SetRank contains three layers, representation layer, encoding layer and ranking layer. The representation layer generates representations of query-document pairs separately with its original learning-to-rank features and ordinal embeddings. The encoding layer jointly processes the documents with multiple sub-layers of multi-head self-attention blocks (MSAB). The final ranking layer calculates the scores and sorts the documents.

\subsubsection{Pre-training Stage}
The pre-training stage is shown in Figure~\ref{fig:SetRankImplementation}(a). During the pre-training stage, the model takes the click log $\mathbb{D}^c$ as the training data, as denoted in Eq.~(\ref{click_data}). $M^{pre}$ also contains an encoder layer $E^{pre}$ and a prediction layer $P^{pre}$.
As for the encoder layer, we use the contextual query and document presentations generate by BERT. Given a query $q^c$ and its associated document set $\mathbf{d}^c= (d_1^c,d_2^c\cdots,d_m^c)$, each of the document $d_i^c$ in $\mathbf{d}^c$ can be represented as a feature vector, the output embedding of the [CLS] token:
$$
\textbf{h}_i^{pre}=E^{pre}([CLS]+q^c+[SEP]+d_i^c+[SEP]).
$$
% \xujun{Representing with BERT?}
all the $(q^c, d_i^c)$ pair in $\mathbf{d}^c$'s representation $\textbf{h}_i^{pre}$ form a matrix $\mathbf{H}^{pre} =\left[\textbf{h}_1^{pre}, \textbf{h}_2^{pre}, \cdots, \textbf{h}_{m}^{pre}\right]$, is then feed into the prediction layer $P^{pre}$, the core architecture of SetRank, several layers of a Multi-head self attention block (MSAB) and a row-wise feed-forward network (rFF) to projects each document representation into one real value as the corresponding click score.

Algorithm $\mathcal{A}^{pre}$ aim to minimize the loss between prediction and click. Due to SetRank is a list ranking approach, list wise loss is used here to optimize the parameters. That is, Eq.~
(\ref{click_loss}) can be written as:
\begin{equation}
\label{equ:lclick_setrank}
\mathcal{L}_{click}= \sum_{(q^c, \mathbf{d}^c, \mathbf{c})\in \mathbb{D}^c} \mathrm{CE_{list}}(P^{pre}(E^{pre}(q^c,\mathbf{d}^c),\mathbf{c}),
\end{equation}
where $\mathrm{CE_{list}}$  denotes the list-wise cross-entropy loss function. The encoder $E^{pre}$ is initialized with a pre-trained BERT model to leverage the pre-trained language model, while the parameters in $P^{pre}$ are learned from the scratch. 

\subsubsection{Fine-tuning Stage}
\begin{figure*}
    \centering
    \includegraphics[width=0.98\textwidth]{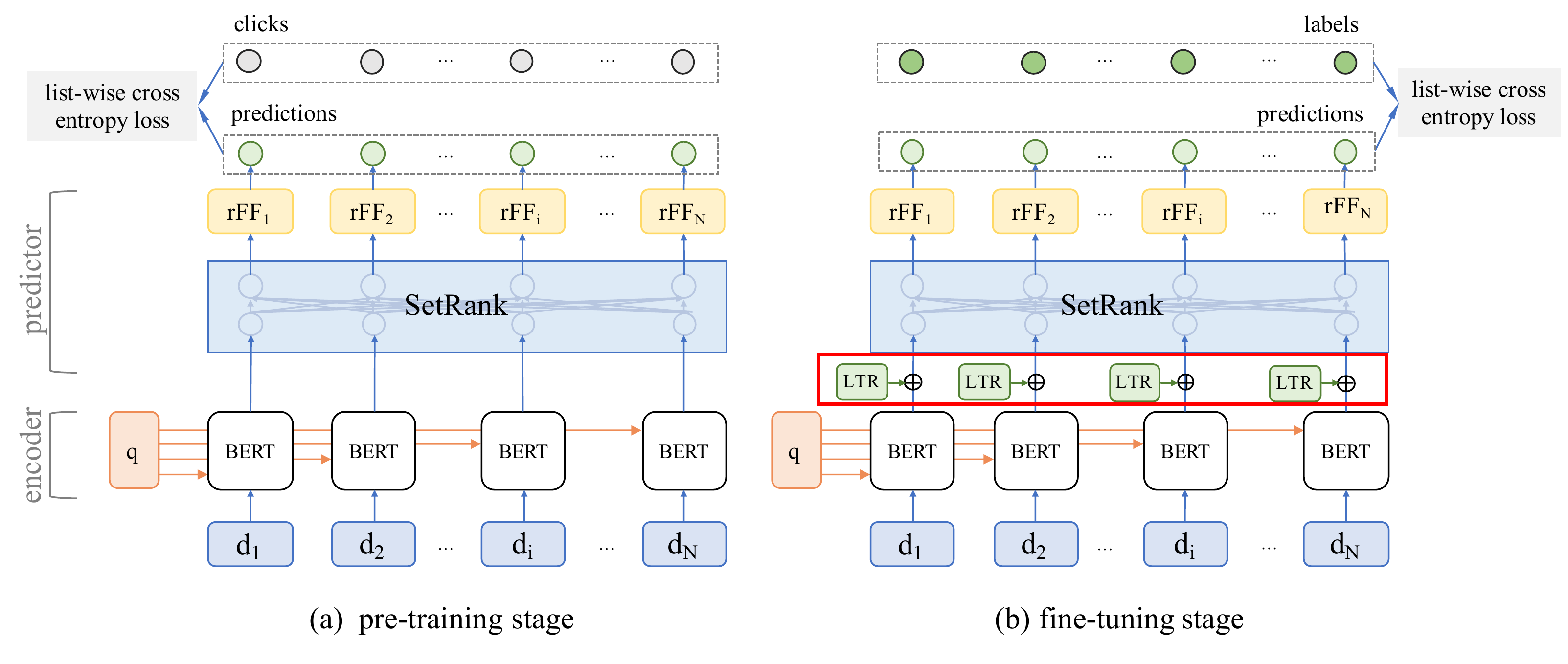}
    \caption{Implementation of Pre-Rank using SetRank as the underlying model. }
    \label{fig:SetRankImplementation}
\end{figure*}

As shown in Figure~\ref{fig:SetRankImplementation}(b), in the fine-tuning stage, for each query $q^l$ a list of documents $\mathbf{d}^l = (d_1^l,d_2^l\cdots,d_m^l)$ is given, as denoted in Eq.~(\ref{label_dataset}).
Similar to that of in the pre-training stage, each $( q^l, d_i^l )$ pair is represent with BERT: 
$$
\textbf{h}^{fine}_{i}=E^{fine}([CLS]+q^l+[SEP]+d_i^l+[SEP]).
$$
All the $(q^l, d^l_i)$ pair in $\textbf{d}^l$'s representation $\textbf{h}_i^{fine}$ form a matrix on the contextual embedding vectors $\textbf{H}^{fine}= \left[\textbf{h}_1^{fine}, \cdots, \textbf{h}_m^{fine}\right]$. 

Meanwhile, the learning-to-rank feature vectors $\psi(q^l,d_i^l)$ for each  $(q^l, d^l_i)$ pair are also respectively generated, forming a feature matrix $\mathbf{\Psi} =[\psi(q^l,d_1^l), \cdots, \psi(q^l,d_m^l)]$. Both $\mathbf{H}^{fine}$ and $\mathbf{\Psi}$ are inputted as the final result of representation layer:
$
\mathbf{X}=[\mathbf{H}^{fine},\mathbf{\Psi}],
$
where $[\cdot,\cdot]$ denotes concatenation of $\mathbf{H}^{fine}$ and learning-to-rank features $\mathbf{\Psi}$. Then $\mathbf{X}$ is feed into SetRank's MSAB blocks and rFF to get the final scores. The ranking can be achieved by sorting the documents according to the scores.

The algorithm $\mathcal{A}^{fine}$ aims to minimize the loss between the list prediction and human label $\mathbf{l}$ of the query-document list:
\begin{equation}
\mathcal{L}_{label}= \sum_{(q^l, \mathbf{d}^l, \mathbf{l})\in \mathbb{D}^l} \mathrm{CE_{list}}\left(P^{fine}\left({[\mathbf{H}^{fine},\mathbf{\Psi}]}\right),\mathbf{l}\right).
\end{equation}

%The fine-tuning model is initialized with the pre-trained SetRank model to leverage the IR knowledge. 

\section{EXPERIMENTS}
We conducted experiments to verify the effectiveness of Pre-Rank on three publicly available IR benchmarks.

\subsection{Experimental Settings}
In this section, we describe the experimental settings, includes datasets, baselines, evaluation metrics, and experimental details.

\subsubsection{Datasets}
In all of the experiments, the ranking models were pre-trained on ORCAS~\cite{craswell2020orcas}, a large-scale publicly available click log corpus. ORCAS contains the real user clicks related to the documents used in the TREC Deep Learning Track. The whole ORCAS corpus contains 1.4 million TREC DL URLs, 18 million connections to 10 million distinct queries after the aggregation and filtering (for satisfying the $k$-anonymity requirements).

% Due to click log can be too revealing of personally or commercially sensitive information, and the noise between query and documents, click logs is not widely use in IR.
%\textbf{Downstream Datasets:}
\begin{table}
\centering
\caption{Statistics of the datasets with human labels. }\label{tab:DatasetStat}
\begin{tabular}{ccccc}  
\toprule
Dataset & Genre      & \#Queries & \#Documents & \#Labeled Q-D pairs  \\ 
\hline
MQ2007  & .gov  & 1,692      & 65,323 & 69,623     \\
MQ2008  & .gov  & 784       & 14,384 & 15,211     \\
TREC19 & web   & 0.37M     & 3.2M   &  388,464    \\
\bottomrule
\end{tabular}
\end{table}

The pre-trained models were then fine-tuned on downstream retrieval datasets with human annotated relevance labels. We conducted fine-tune on three datasets MQ2007, MQ2008, and TREC2019, which are published in LETOR 4.0~\cite{qin2010letor} and TREC19~\cite{craswell2020overview}. 
Table~\ref{tab:DatasetStat} lists some statistics of the three datasets.
% More specifically, the LETOR 4.0 contains two separate datasets: MQ2007 and MQ2008 respectively. 
MQ2007 contains 1,692 queries 65,323 documents and 69,623 expert-labeled query-document pairs. MQ2008 contains 784 queries 14,384 documents and 15,211  expert-labeled query-document pairs. The number of queries in MQ2008 is relatively small, making it insufficient to learn deep learning model. In the experiments, we followed the practices in~\cite{pang2017deeprank} and combined the training set of MQ2007 with that of MQ2008 while keeping the validation and test sets unchanged. We still denoted the new dataset as MQ2008. MQ2007 and MQ2008 in total contain 69,623 and 84,834 query documents pairs. The two datasets consist of not only the raw query and document texts, but also 46 dimensions handcrafted features for each query-document pair.
TREC 2019 Deep Learning Track benchmark is a large-scale ad-hoc retrieval dataset, which served two tasks: document retrieval and passage retrieval. Both tasks have large training sets with human relevance assessments, derived from MS MARCO~\cite{bajaj2016ms}. The document retrieval task has a corpus of 3.2 million documents with 367 thousand training queries, and 388,464 labeled query-document pairs. Please note TREC19 only provides the relevant documents for each query.  In this paper, we conducted experiments on the sub-task of ``top-100 reranking'' in document retrieval.   
% \begin{itemize}
% \item {\textbf{Million Query Track 2007 (MQ2007)}}: 
% \item {\textbf{Million Query Track 2008 (MQ2008)}}: is another LETOR benchmark dataset with 784 queries, which also leverages the Gov2 Web collection.
% \end{itemize}
\subsubsection{Baselines and Evaluation Metrics}
Several types of relevance ranking baselines, including the traditional relevance ranking models, learning-to-rank models, the state-of-the-art neural IR models, were selected in the experiments for comparisons: 
%\liang{Why these baselines? Using system view and user view? I think all of these baselines are system view? The pre-trained model by our Pre-Rank is user view?}
\begin{itemize}
\item\textbf{BM25}~\cite{robertson1994some}: a classical and widely used model for relevance ranking; 
\item\textbf{AdaRank}~\cite{xu2007adarank}: a traditional learning-to-rank model that aims to directly optimize the performance measure based on boosting;
\item\textbf{LambdaMart}~\cite{burges2010ranknet}: a widely used traditional listwise learning-to-rank model based on gradient boosting;
% \item\textbf{DeepRank}~\cite{pang2017deeprank}: a neural IR that models relevance by simulating the human judgement process;
% \item\textbf{HINT}~\cite{fan2018modeling}: a neural IR model that models the relevance signals at different granularities to compete with each other, for generating the final relevance assessment;
\item\textbf{BERT}~\cite{devlin2018bert}: a state-of-the-art pre-trained language model which uses MLM and next sentence prediction (NSP) to pre-train the contextual language representation and sentence pair representation. The pre-trained representations can be fine-tuned to a number of downstream tasks, including relevance ranking.  
\item\textbf{SetRank}~\cite{pang2020setrank}: a permutation-invariant neural IR model which has the ability to model cross-document interactions so as to capture local context information under a query;
\item\textbf{PROP}\cite{ma2021prop}: a recently proposed pre-trained IR model that tailored the training object during pre-training. In practice, PROP was pre-trained on Wikipedia and MS MARCO Document Ranking dataset, denoted as PROP$_{wiki}$ and PROP$_{MARCO}$, respectively. 
\end{itemize}

%\subsubsection{Evaluation Measures}
To evaluate the performances of Pre-Rank and the baselines on MQ2007 and MQ2008, we followed the original data partitions provided by LETRO4.0 and conducted 5-fold cross validation. The average results were reported. As for the evaluation measures, we used the Precision at 10 (P@10), Normalized Discounted Cumulative Gain at 10 (NDCG@10), and Mean Average Precision (MAP) ~\cite{robertson2000evaluation}. 

To evaluate TREC19, we tested the performances of the proposed Pre-Rank and baselines on the dev set. As for the evaluation measures, we used Recall at 10 (Recall@10) and Mean Reciprocal Rank at 10 (MRR@10).

\subsubsection{Experimental Details}
During the pre-training procedures, to incorporate relevance assessment from the real users into the ranking model, we performed the pre-training on the ORCAS dataset. ORCAS collected large-scale real user queries, documents, and clicks from a search engine. Like the pre-training of BERT, we also performed two tasks on ORCAS dataset, namely mask language model and click prediction. The procedure of the mask language model task is identical to that of the original BERT model. We randomly masked 15\% tokens and made the model to ``restore'' them. As for the click prediction, the positive instances are derived from the click data of ORCAS, and the negative instances are randomly selected from the top 100 documents of the corresponding query if users did not click them. %Though above two tasks, we achieve Mask Language Model and click pre-trained models. 
We combined the two tasks for training the pre-trained model by first running the mask language model task and then running the click prediction task.

During the fine-tuning procedures, we used the pre-trained models to re-rank the candidate documents provided by the datasets.The fine-tuning made use of the raw text of queries, documents, and the handcrafted features provided in the dataset.
For MQ2007 and MQ2008, the learning rate was tuned between $10^{-5}$ to $2\times 10^{-5}$. For fine-tuning Pre-Rank(BERT), we randomly selected 20 documents from the datasets for each query. The input was truncated to 256. For fine-tuning Pre-Ran(SetRank), the list sizes were set to 20 and the input of raw texts were also truncated to 256 for fair comparisons. When conducting the tests on the fine-tuned model, all documents are inputted to get the final ranks. For TREC19 dataset, we randomly select 40 documents per query from the top100 retrieval results officially provided by this dataset, and other settings as the same as MQ2007 and MQ2008 datasets.

Pre-Rank utilized handcrafted relevance features at the fine-tuning stage. These features are combined with the pre-trained representations for making accurate relevance predictions. On MQ2007 and MQ2008, we directly used the learning-to-rank features provided by the LETOR4.0 corpus. Each query-document pair is represented as a 46-dimensional real vector. On TREC19, we extracted 21-dimensional features for each query-document pair. These features reflect the information of the query-level, document-level, and the interaction between the query and document, including query length, BM25 score, tf-idf, query-document matching with bi-grams, tri-grams similarity, word2vec similarity, etc. 

For the traditional IR models and learning-to-rank models (i.e., BM25, AdaRank, and LambdaMart), we used the implementations shared in ~\cite{pang2017deeprank}. The implementations of BERT and Pre-Rank are based on the popular Transformers library~\footnote[1]{\url{https://github.com/huggingface/transformers}}. As for BERT, the encoder layer of Pre-Rank(BERT) and the encoder layer of Pre-Rank(SetRank) are initialized by BERT$_{\textrm{BASE}}$'s checkpoint release by Google~\footnote[2]{\url{https://github.com/google-research/bert}}.
% The source code and experiments can be found at \url{http://github.com/hide_for_anonymous_review}

\subsection{Experimental results }

\begin{table*}
\centering
\caption{Ranking performance comparisons between Pre-Rank and the baselines on three benchmarks with human labels. `*' and `$\dagger$' respectively indicate the improvements over BERT and SetRank are statistically significant ($p$-value$< 0.05$).}
\label{tab:exp_main}
\begin{tabular}{c|ccc|ccc|cc} 
\toprule
\multirow{2}{*}{Model} & \multicolumn{3}{c|}{MQ2007}                      & \multicolumn{3}{c|}{MQ2008}                       & \multicolumn{2}{c}{TREC19}       \\
                       & P@10          & NDCG@10        & MAP            & P@10           & NDCG@10        & MAP            & MRR@10         & Recall@10       \\ 
\hline\hline
BM25                   & 0.366         & 0.414          & 0.450          & 0.245          & 0.220          & 0.465          & 0.234          & 0.473           \\
AdaRank                & 0.373         & 0.439          & 0.460          & 0.247          & 0.222          & 0.468          & 0.271          & 0.533           \\
LambdaMart             & 0.384         & 0.446          & 0.468          & 0.251          & 0.231          & 0.478          & 0.273          & 0.529           \\
BERT                   & 0.418         & 0.495          & 0.500          & 0.252          & 0.247          & 0.502          & 0.370          & 0.632           \\
SetRank                & 0.418         & 0.497          & 0.498          & 0.255          & 0.249          & 0.498          & 0.383          & 0.638           \\
PROP$_{wiki}$               & 0.432         & 0.523          & -              & 0.267          & 0.262          & -              & 0.360          & 0.622           \\
PROP$_{MARCO}$              & 0.430         & 0.522          & -              & \textbf{0.269} & \textbf{0.266} & -              & 0.360          & 0.628           \\ 
\hline
Pre-Rank(BERT)         & 0.430$^{*}$     & 0.520$^{*}$      & 0.521$^{*}$      & 0.255$^{*}$      & 0.252$^{*}$      & 0.514$^{*}$      & 0.376$^{*}$      & 0.644$^{*}$       \\
Pre-Rank(SetRank)      & \textbf{0436}$^{\dagger}$ & \textbf{0.526}$^{\dagger}$ & \textbf{0.525}$^{\dagger}$ & 0.258$^{\dagger}$          & 0.258$^{\dagger}$          & \textbf{0.521}$^{\dagger}$ & \textbf{0.388}$^{\dagger}$ & \textbf{0.648}$^{\dagger}$  \\
\bottomrule
\end{tabular}
\end{table*}

Table~\ref{tab:exp_main} reports the experimental results of the proposed Pre-Rank and all of the baselines on the downstream datasets of MQ2007, MQ2008, and TREC19. The baselines' results mainly follow~\cite{pang2017deeprank}. For PROP$_{wiki}$ and PROP$_{MARCO}$ we used the numbers reported in~\cite{ma2021prop} thus have no results on TREC19 and in terms of MAP on MQ2007 and MQ2008. The results on TREC19 of PROP are based on the released models in~\cite{ma2021prop}.  

By comparing Pre-Rank (BERT) with BERT and Pre-Rank (SetRank) with SetRank, we can see that Pre-Rank outperformed the underlying neural IR model of BERT and SetRank with large margin. We conducted significant tests on the improvements. The results indicate that all of the improvements over the raw underlying ranking models are significant ($p$-value $< 0.05$). Considering that BERT and SetRank are already very strong neural IR baselines, the results indicate the effectiveness of the Pre-Rank framework. It also verified the effectiveness of Pre-Rank's approach that involving both the user's view and the system's view of relevance signals in one model using the pre-training and fine-tuning paradigm. Comparing the two implementations, we found that Pre-Rank (SetRank) performed better than Pre-Rank (BERT), verified the advantages of the permutation-invariant ranking model.   

From the results, we also found that: (1) the neural IR models of BERT and SetRank can obtain better results than traditional models such as BM25, AdaRank and LambdaMART, indicating that automatically learned features can capture more relevance signals other than the handcrafted learning-to-rank features; 

(2) the pre-trained ranking model PROP~\cite{ma2021prop} improved BERT on MQ2007 and MQ2008 by a huge margin, verified the power of the pre-training and fine-tuning paradigm;
(3) the proposed Pre-Rank (SetRank) outperformed PROP in most cases, verified the effectiveness and necessity of modeling the relevance from both user's view and system's view. 

\subsection{Discussions}
We conducted experiments to show the reasons that our approaches outperformed the baselines and impacts of different parameter settings. 

\subsubsection{Impact of handcrafted features at the fine-tuning stage}
One of the unique characteristics of Pre-Rank is its ability to involve both the pre-trained representations and handcrafted learning-to-rank features. In the experiments, we tested the effects of the handcrafted features. Specifically, we tested the ranking accuracies of Pre-Rank (BERT) and Pre-Rank (SetRank) when the wide parts (which are responsible for injecting learning-to-rank features) were removed during the fine-tuning. In the experiments, all of the inputs were under the same setting for each dataset. Table~\ref{Table_LTR_features} lists the experimental results. In the table, the Pre-Rank (BERT) and Pre-Rank (SetRank) versions with and without the handcrafted features were denoted as ``/w ltrFtr'' and ``w/o ltrFtr''. From the results, we can see that in most cases combined with the automatically pre-trained features, the handcrafted features can further improve the ranking accuracy under the Pre-Rank framework. The results verified that though the powerfulness of the pre-trained representations, the expert knowledge encoded in the traditional learning-to-rank features (one aspect of the system's view on relevance) is still valuable for relevance ranking.  

\begin{table*}
\centering
\caption{Impacts of the handcrafted learning-to-rank features at the fine-tuningstage. }
\label{Table_LTR_features}
\begin{tabular}{cc|ccc|ccc|cc} 
\toprule
                                     &             & \multicolumn{3}{c|}{MQ2007} & \multicolumn{3}{c|}{MQ2008} & \multicolumn{2}{c}{TREC19}  \\
                                     &  Settings  & P@10 & NDCG@10 & MAP       & P@10 & NDCG@10 & MAP       & MRR@10 &  Recall@10                  \\ 
\hline\hline
\multirow{2}{*}{Pre-Rank (BERT)}    & w/o LtrFtr     & 0.430 & 0.516 & 0.518 &0.256 &0.252    &0.513      & 0.375  &      0.642       \\
                                     & w/ LtrFtr      & 0.430 & 0.520 & 0.521 &0.255 &0.252    &0.514      & 0.376 & 0.644      \\
\hline
\multirow{2}{*}{Pre-Rank (SetRank)} & w/o LtrFtr     & 0.433 & 0.521 & 0.520 &0.256 &0.252    &0.510      & 0.389  &  0.643          \\
                                     & w/ LtrFtr      & 0.436 & 0.526 & 0.525 &0.258 &0.258    &0.521      & 0.388  &     0.648               \\
\bottomrule
\end{tabular}
\end{table*}

\subsubsection{Selection of the learning objectives at the pre-training stage}
During the pre-training stage, Pre-Rank utilized two learning objectives: the traditional mask language model (MLM) objective based on the raw texts, and the click prediction objective based on the real user's click activities. We conducted experiments to investigate the impacts of these two learning objectives in Pre-Rank. Specifically, we removed one of the objectives and pre-trained the initial representations with the remained objective. The fine-tuning stage is kept unchanged. Table~\ref{table_pretrain_objs} shows the performances of the variations of Pre-Rank (BERT) and Pre-Rank (SetRank) in which the pre-training objectives are adjusted. In the table, we denote the experiments pre-training with MLM objective only, pre-training with click prediction objective only, and pre-training with MLM and click prediction as ``w/MLM'', ``w/click'', and ``w/MLM + Click'', respectively. From the results reported in Table~\ref{table_pretrain_objs}, we found that Pre-Rank models pre-trained with click prediction objective only (``w/Click'') performed similarly to the full version: pre-training with both MLM and click prediction objectives (``w/MLM + Click''). On the other hand, the model that pre-trained with MLM objective only (``w/MLM'') performed much worse. The phenomenon can be observed in both the experiments with Pre-Rank (BERT) and Pre-Rank (SetRank). The results clearly indicate that: (1) the activities from the real users (user's view of relevance) are critical relevance signals for relevance ranking; (2) through existing studies have proven MLM to be an effective pre-training objective for many NLP related downstream tasks, it seems that MLM is not an optimal selection for relevance ranking. Both the user's view and the system's view on relevance also rarely touch the NLP knowledge of the texts in the queries and documents, especially the knowledge that can be derived from the mask language modeling. Therefore, it is easy to understand why very limited improvements can be observed after adding the MLM objective to the click prediction objective. 

\begin{table*}
\centering
\caption{Impacts of the pre-training objectives in Pre-Rank. }
\label{table_pretrain_objs}
\begin{tabular}{cc|ccc|ccc|cc} 
\toprule
                                     &             & \multicolumn{3}{c|}{MQ2007} & \multicolumn{3}{c|}{MQ2008} & \multicolumn{2}{c}{TREC19}  \\
                                     & Settings            & P@10 & NDCG@10 & MAP       & P@10 & NDCG@10 & MAP       & MRR@10 &  Recall@10                  \\ 
\hline\hline
\multirow{4}{*}{Pre-Rank(BERT)}      %& w/BERT base       & 0.418 & 0.495    &0.500      &0.247 & 0.252    &0.502      &  0.370      &    0.632        \\
                                    & w/MLM       &0.424 &0.502    &0.504      &0.253 &0.246    &0.502      &  0.371  &   0.636            \\
                                     & w/MLM+Click &0.430 &0.520    &0.521      &0.255 &0.252    &0.514      &    0.376    &    0.644         \\
                                     & w/Click     &0.432 &0.521    &0.518      &0.255 &0.252    &0.507      &   0.373  &     0.641         \\ 
\hline
\multirow{4}{*}{Pre-Rank(SetRank)}  % & w/BERT base       &0.418 &0.497    &0.498      &0.249 &0.255    &0.498      &   0.383   &   0.638      \\
                                    & w/MLM       &0.418 &0.487    &0.496      &0.256 &0.253    &0.515      &   0.387   &   0.644      \\
                                     & w/MLM+Click &0.436 &0.526    &0.525      &0.258 &0.258    &0.521      &   0.388    &    0.648           \\
                                     & w/Click     &0.436 &0.526    &0.525      &0.259 &0.258    &0.523      &   0.387    &    0.645          \\
                                     
\bottomrule
\end{tabular}
\end{table*}

\subsubsection{Ability to improve long queries}
\begin{figure}
    \centering
    \includegraphics[width=0.5\textwidth]{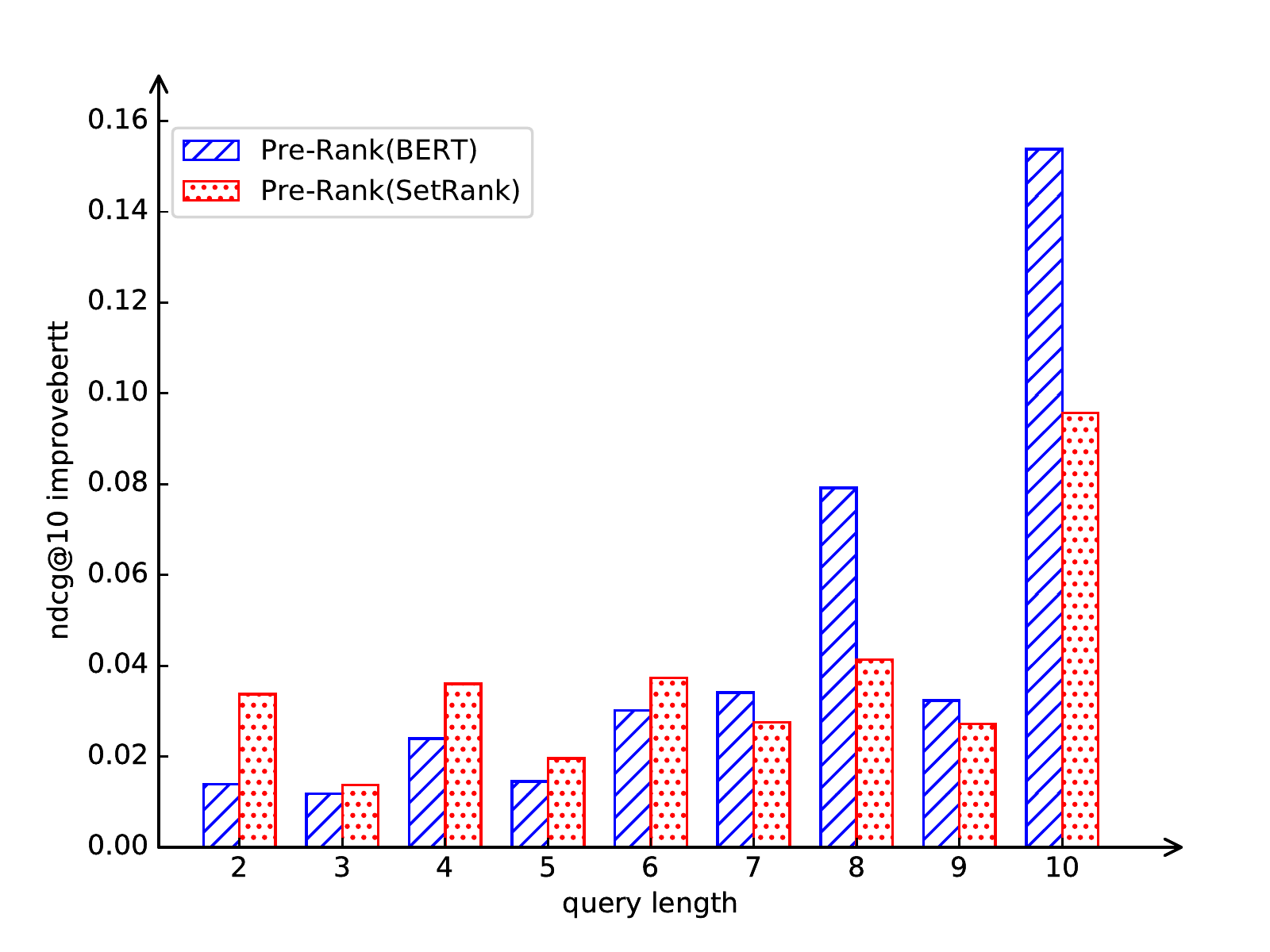}
     \caption{NDCG@10 improvements of Pre-Rank (BERT) and Pre-Rank(SetRank) over the underlying ranking models w.r.t. different query lengths on MQ2007.  }
    \label{fig:compare}
\end{figure}

We also conducted experiments to show on which kinds of queries our approaches can perform well. Specifically, we categorized the queries of MQ2007 into different query groups according to the lengths of the queries. For each query group, we tested the performance improvements of Pre-Rank (BERT) over the underlying ranking model of BERT, in terms of NDCG@10, and the performance improvements of Pre-Rank (SetRank) over the underlying ranking model of SetRank in terms of NDCG@10. Figure~\ref{fig:compare} showed the NDCG@10 improvements w.r.t. the query groups. The results are the average values over the 5 folds' test set. We can see that there is a trend that Pre-Rank can achieve more improvements on the query groups with long query lengths (e.g., the group whose query length is 10). This is particularly obvious in the case of Pre-Rank (BERT). One reason for the phenomenon is that by pre-training on raw text from the user's view, Pre-Rank models more  semantic relevance and therefore improves when the query is longer and requires more semantic information.

\subsubsection{Effects of combining relevance signals from user's view and system's view} 
\begin{figure}
    \centering
    \includegraphics[width=0.5\textwidth]{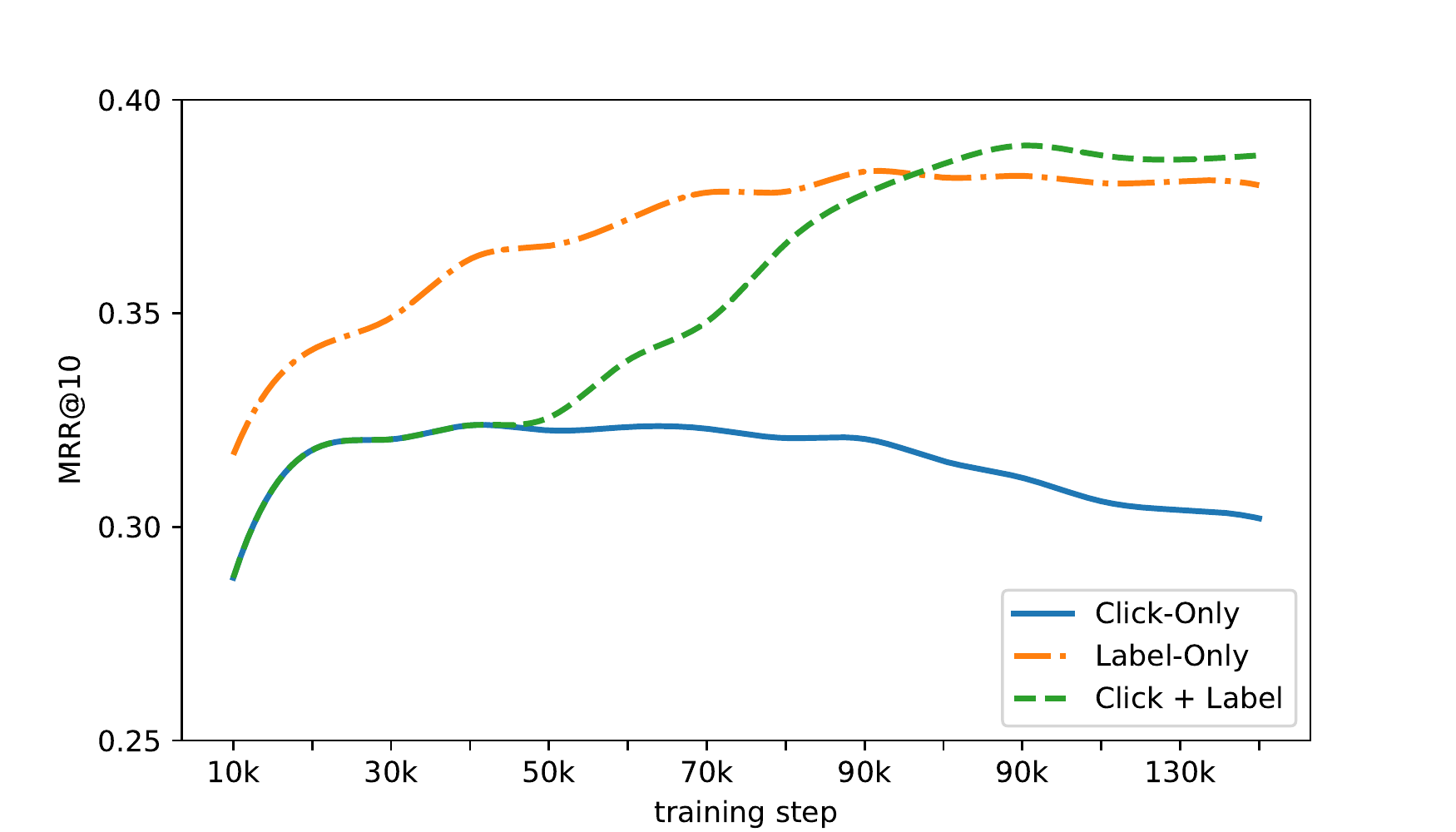}
    \caption{Performance curves of SetRank models when respectively trained with user clicks (ORCAS), with labeled data (TREC19), and first with user clicks and then with labeled data. }
    \label{fig:EffectsCombineUserSystem}
\end{figure}

 We conducted experiments to test the effects of combining the user's view and the system's view on relevance in learning ranking models. Specifically, we trained SetRank models based on three settings of the training data: (1) trained on the ORCAS dataset that only contains clicks from real users (denoted as ``Click-Only''); (2) trained on the TREC19 dataset that only contains relevance labels by the experts (denoted as ``Label-Only''); (3) first trained on the ORCAS click log, and then continued the training on TREC19 expert-labeled dataset (denoted as ``Click + Label''). We tested the performances on TREC19's dev set and Figure~\ref{fig:EffectsCombineUserSystem} shows the performance curves of these three settings w.r.t. training steps in terms of MRR@10.
%  \footnote{Note that MRR is the evaluation measure on TREC19 recommended by the official instructions.}. 
 From the results, we can conclude that (1) the large-scale click log (relevance signals from the user's view) is effective in training ranking models in the early stages. However, as the training goes on, the noise and biases in the click log hurt the training procedure and hinder the further improvements; (2) the limited but clean labeled data (relevance signals from the system's view) is useful in training and the model's performance improved steadily until convergence; (3) the ranking performances can be further improved when the model trained on ORCAS was continually trained on TREC19 data. This is because these two views are complementary: the clean and unbiased labeled data in TREC19 relieves the noise and bias issues with the click log, while the large-scale click log provides massive relevance signals to overcome the scale limitation. 

\section{CONCLUSION}
This paper proposes a novel relevance ranking framework under the pre-training and fine-tuning paradigm, for modeling the relevance information from both the user's view and the system's view. The framework, called Pre-Rank, first pre-trains the initial representations on real user's search activities (e.g., click log) and then fine-tunes the ranking model on expert-labeled data with handcrafted learning-to-rank features. We implemented two versions of Pre-Rank which used BERT and SetRank as its underlying ranking model, respectively. Experimental results on publicly available benchmarks showed that after pre-training on large-scale user activities data, both of the pre-trained Pre-Rank versions can be well fine-tuned on three benchmarks and significantly enhanced the ranking performances. 

\begin{acks}
This work was funded by the National Key R\&D Program of China (2019YFE0198200), the National Natural Science Foundation of China (No. 61872338, No. 61906180, No. 61832017), Beijing Academy of Artificial Intelligence (BAAI2019ZD0305), and Beijing Outstanding Young Scientist Program NO. BJJWZYJH012019100020098.
\end{acks}

\balance
\bibliographystyle{ACM-Reference-Format}
\bibliography{sample-base}

\end{document}